%% file: main.tex
\documentclass{emulateapj}

\newcommand{\Kepler}{{\it Kepler}}

\newcommand{\Alopeke}{`$\!$Alopeke}

\newcommand{\be}{\begin{equation}}
\newcommand{\ee}{\end{equation}}

\newcommand{\meh}{[M/H]}

\newcommand{\rearth}{R$_\oplus$}

\newcommand{\thisstarone}{EPIC~246151543}
\newcommand{\thisstartwo}{EPIC~246078672}
\newcommand{\thisplanetone}{EPIC~246151543~b}
\newcommand{\thisplanettwo}{EPIC~246078672~b}

\usepackage{xcolor}
\usepackage{hyperref}
\usepackage{breakurl}
\usepackage{amsmath}
\usepackage{graphicx}
\usepackage{verbatim}
\usepackage{booktabs}
\usepackage{textcomp}   

\newcommand{\ron}{}

\slugcomment{}

\shorttitle{Identifying Exoplanets with Deep Learning in K2 Data}
\shortauthors{Dattilo et al.}

\begin{document}


\title{Identifying Exoplanets with Deep Learning II: Two New Super-Earths Uncovered by a Neural Network in K2 Data}
\author{Anne Dattilo\altaffilmark{1,$\dagger$}, Andrew Vanderburg\altaffilmark{1,$\star$}, Christopher J. Shallue\altaffilmark{2}, Andrew W. Mayo\altaffilmark{3,$\ddagger$}, Perry Berlind\altaffilmark{4}, Allyson Bieryla\altaffilmark{4}, Michael L. Calkins\altaffilmark{4}, Gilbert A. Esquerdo\altaffilmark{4}, Mark E. Everett\altaffilmark{5}, Steve B. Howell\altaffilmark{6}, David W. Latham\altaffilmark{4}, Nicholas J. Scott\altaffilmark{6}, Liang Yu\altaffilmark{7}}

\altaffiltext{1}{Department of Astronomy, The University of Texas at Austin, Austin, TX 78712, USA}
\altaffiltext{2}{Google Brain, 1600 Amphitheatre Parkway, Mountain View, CA 94043}
\altaffiltext{3}{Astronomy Department, University of California, Berkeley, CA 94720, USA}
\altaffiltext{4}{Harvard--Smithsonian Center for Astrophysics, 60 Garden St., Cambridge, MA 02138, USA}
\altaffiltext{5}{National Optical Astronomy Observatory, 950 North Cherry Avenue, Tucson, AZ 85719}
\altaffiltext{6}{Space Science and Astrobiology Division, NASA Ames Research Center, Moffett Field, CA 94035}
\altaffiltext{7}{Department of Physics and Kavli Institute for Astrophysics and Space Research, Massachusetts Institute of Technology, Cambridge, MA 02139, USA}

\altaffiltext{$\star$}{NASA Sagan Fellow}
\altaffiltext{$\ddagger$}{NSF Graduate Research Fellow}
\altaffiltext{$\dagger$}{\url{anne.dattilo@utexas.edu}}


\begin{abstract}

{\ron For years, scientists have used data from NASA's \Kepler\ Space Telescope to look for and discover thousands of transiting exoplanets. In its extended K2 mission, \Kepler\ observed stars in various regions of sky all across the ecliptic plane, and therefore in different galactic environments.} Astronomers want to learn how the population of exoplanets are different in these different environments. {\ron However, this requires an automatic and unbiased way to identify the exoplanets in these regions and rule out false positive signals that mimic transiting planet signals.} We present a method for classifying these exoplanet signals using deep learning, a class of machine learning algorithms that have become popular in fields ranging from medical science to linguistics. We modified a neural network previously used to identify exoplanets in the \Kepler\ field to be able to identify exoplanets in different K2 campaigns, which range in galactic environments. We train a convolutional neural network, called \texttt{AstroNet-K2}, to predict whether a given possible exoplanet signal is really caused by an exoplanet or a false positive. \texttt{AstroNet-K2} is highly successful at {\ron classifying exoplanets and false positives}, with accuracy of 98\% on our test set. {\ron It is especially efficient at identifying and culling false positives, but for now, still needs human supervision to create a complete and reliable planet candidate sample.} We use \texttt{AstroNet-K2} to identify and validate two previously unknown exoplanets. Our method is a step towards automatically identifying new exoplanets in K2 data and learning how exoplanet populations depend on their galactic birthplace.

\end{abstract}


\section{Introduction}

In 2013, NASA\rq s \Kepler\ Space Telescope suffered the failure of the second of its four reaction wheels used to orient and stabilize the spacecraft. {\ron Before the mechanical failure, \Kepler\ was pointed at a single field near Cygnus in order to obtain multi-year photometry that would enable detection of long-period planets}, but the failure rendered \Kepler\ unable to point stably at its original field of view. Though \Kepler\ could no longer observe its original field, a clever engineering solution was devised in which \Kepler\ observed different fields across the ecliptic plane, with worsened pointing precision, for periods of about 80 days \citep{howell}. {\ron This new observing mode, called K2, enabled observations capable of detecting exoplanets, but the analysis of this data would be more difficult than before}; K2 data are marred by systematic noise from the telescope's now unstable pointing, and the shorter observational baseline meant less data was collected for each star. However, as data analysis techniques were developed to remove the large systematic errors from K2 data \citep[e.g.][]{vj14}, K2 delivered on its promise of continuing to detect small transiting exoplanets in fields across the ecliptic plane \citep[e.g.][]{crossfield16, mayo, livingston1, livingston2}. 

The fact that, unlike the original \Kepler\ mission,  K2 observed fields across the ecliptic plane presents a powerful opportunity. \Kepler\ revolutionized our knowledge of the occurrence rates of small exoplanets \citep{burke, dc15}, but \Kepler's discoveries come from only one part of the sky. Because K2 looked at different regions of the sky, it observed stars and planets that formed in different galactic environments. We know that stars in different parts of the sky and different birth environments have different properties \citep{west2008, boeche2013}, so it is reasonable to expect that planets in these different environments might have different properties as well. {\ron By observing fields across the ecliptic, K2 can study stars both near the galactic plane, where many young stars can be found, and the older stars a few scale heights away.} K2 also has observed (and discovered planets in) several nearby open clusters and associations \citep{obermeier, mann16, rizzuto18, livingstonhyades, v18}. K2's observational breadth could enable determining occurrence rates for planets around different stellar populations and in different galactic environments, a key to understanding how and why planets form.

So far, however, K2 data have not yet yielded many measurements of planet occurrence rates \citep[with a few notable exceptions involving relatively small numbers of unusual host stars,][]{vans, demory}. One reason planet occurrence rates have not yet been widely measured with K2 data is the lack of an accurate automated system for identifying which signals detected by the spacecraft are likely to be genuine exoplanets. During the original \Kepler\ mission, this function was largely performed by the \Kepler\ team's Robovetter system \citep{thompsonlpp, coughlin, thompson2018}, a decision tree designed to mimic the steps taken by a human to decide whether any given signal was likely planetary. Systems like the Robovetter decouple planet occurrence rates from human judgment, biases, and non-uniformity, and make it possible to measure quantities like the false positive and false negative rates of a given catalog. Once these rates are known, astronomers can correct the resulting occurrence rates for these errors. A rapid, automatic, and characterizable method to identify planet candidates in K2 data will enable occurrence rate studies, which could reveal the differences in exoplanet populations in different galactic environments.  

In this work, we address the task of creating an automatic system for identifying planet candidates in K2 data using \textit{deep learning}, a modern machine learning technique.  Recently, various machine learning algorithms have started to be used for this and other related purposes. A pioneering machine learning algorithm for classifying planet candidates and false positives in \Kepler\ data was the Autovetter \citep{autovetter}, a random forest classifier that made decisions based on metrics generated by the \Kepler\ pipeline. More recently, \citet{millholland} used a logistic regression algorithm to attempt to detect \textit{non-transiting} hot Jupiters, and \citet{armstrong2017} used self-organizing maps and random forests to classify variability, eclipsing binaries, and transiting planets in both Kepler/K2 data and data from ground-based surveys \citep{armstrongngts, schanche}. Neural networks, a type of deep learning model, have been used successfully for detecting transits in simulated \Kepler-like data \citep{zucker, pearson} and for classifying planet candidates and false positives in \Kepler\ data \citep{shallue, ansdell}.



In this paper we expand on the work of \citet{shallue}, which was designed to distinguish planet candidates and false positives in \Kepler\ data, to classify these signals in data from the K2 mission. Following \citet{shallue}, we use a supervised convolutional neural network architecture, but make several key modifications to enhance the network's ability to classify signals in the qualitatively different K2 dataset. Our paper is organized as follows: In Section~\ref{training set}, we discuss the creation of the training set, including the source of the data used in the representations sent to the neural network, and the method by which we labeled the training, test, and validation sets used in our work. Section~\ref{nn model} describes the architecture and training of our neural network, which we call \texttt{AstroNet-K2}. In Section~\ref{eval}, we evaluate the success of our neural network in classifying planet candidates and false positives. We proceed to test our neural network's ability to classify previously un-labeled signals in Section~\ref{new}, where we detect a handful of new planet candidates and describe the steps we took to statistically validate two of these candidates as \textit{bona fide} exoplanets. We discuss our results in Section~\ref{discussion}, including the prospects for using a similar method for determining exoplanet occurrence rates in K2 data, and we give avenues for future improvement of our method. Finally, we conclude in Section~\ref{summary}. 


\section{Training Set}\label{training set}

Our neural network utilizes ``supervised learning,'' which means we provide the neural network with a labeled set of examples, from which it can learn. We call this a \textit{training set}.

In this paper, we take advantage of work our team has done over the past four years as K2 data has been released to the public. Over that time, our team has routinely produced light curves, searched for transits, and identified likely planet candidates in support of a wide variety of scientific objectives \citep{mayo, rodriguez2018}. Here, we give a brief overview of how we produced these light curves and training set examples, much of which follows the methods outlined by \citet{v16}. We also describe some additional steps we took to refine the dataset for use as a neural network training set (Section~\ref{vetting} and beyond). 

\subsection{Identifying Threshold Crossing Events}\label{generation}

{\ron Our training set consists of possible planet signals that, following the naming convention in the literature, we refer to as ``Threshold Crossing Events'' or TCEs.} These are potentially periodic signals (decreases in the brightness of a star) that have been detected by an algorithm designed to search for transiting exoplanets in a light curve. A TCE is characterized by the star on which it is observed, the period on which the signal appears to repeatedly cause the star to dim, the time at which the first repetition of the dimming signal is observed, and the duration of the dimming signal. 

\subsubsection{Light curve production}\label{lc}
The first step to identify TCEs is to produce light curves that can be searched. In this work, we used light curves that one of us (AV) has produced since 2014 using methods described by \citet{vj14} and \citet{v16}, which we outline here in brief. Upon the public release of data from each K2 campaign, we downloaded the K2 target pixel files from the Mikulski Archive for Space Telescopes (MAST) and extracted raw light curves by summing the flux contained within in 20 different stationary photometric apertures at each time stamp. Due to the \Kepler\ spacecraft's unstable pointing in its K2 operating mode, these raw light curves exhibit large systematic features that impede the detection of planetary transits. We corrected for these systematics by decorrelating the systematic variability from the spacecraft's motion. The resulting light curves from each of the 20 different photometric apertures are publicly available in the K2SFF High Level Science Product\footnote{\url{https://archive.stsci.edu/prepds/k2sff/}}, hosted at the MAST. We then selected an ``optimal'' light curve by determining which photometric aperture produced the high-pass-filtered light curve with the lowest photometric scatter on timescales shorter than about 6 hours. The optimal light curves used in this work in some cases are different from the optimal light curves chosen for the public K2SFF light curves, as the K2SFF optimal light curves were chosen by minimizing photometric scatter in light curves that had not been high-pass-filtered. 

\subsubsection{Transit Search}\label{search}
After producing systematics-corrected light curves and selecting optimal photometric apertures for each K2 target, we searched the targets for periodic dipping signals as described by \citet{v16}. We performed the transit search by calculating a Box Least Squares (BLS) periodogram \citep{kovacs}, identifying and recording signals exceeding our signal-to-noise threshold of $S/N=9$, removing the signals, and re-calculating the BLS peridodogram until no significant signals remain in each light curve. We call each signal stronger than our $S/N$ threshold a ``Threshold Crossing Event'', or TCE. For most stars, our search does not identify any periodicities strong enough to be considered TCEs, but some stars have multiple TCEs. {\ron These TCEs can be triggered by only 1 (or more) transit event(s). The BLS was performed on campaigns 0-16 and resulted in a total of 51,711 TCEs.}


\subsection{Labeling Threshold Crossing Events}
{\ron The transit search resulted in a total of 51,711 identified TCEs. We categorized a majority of them (31,575) by hand in order to produce the labeled training set for our neural network.} Our end goal was to sort the TCEs into two categories: planet candidates and false positives. We used a two-step process to label the TCEs. We started with triage, a necessarily quick process to go through large numbers of TCEs, and then gave more careful consideration to the subset that most resembled possible transit signals through vetting. This is a time-efficient process that we believe results in majority accurate classifications (more than half of TCEs that were not discarded ended up being classified as planet candidates). 

\subsubsection{Triage}\label{triage}

We performed triage on the majority of the TCEs detected by our transit search. Early on in the K2 mission (in particular Campaigns 0-3, \citealt{v16}), we scrutinized every single TCE returned by our transit search. In Campaign 4 and beyond, however, we found that it was more efficient to only examine the first TCE detected around each star unless the first TCE was itself classified as a planet candidate. Because the first TCE detected around a star is the strongest signal present, in most cases, when the first TCE is either an astrophysical false positive or an instrumental artifact, all other TCEs around that star are also likely false positives. This choice saved us the effort of performing triage on approximately 20,000 additional TCEs, at the expense of a handful of missed exoplanets. 

\begin{figure*}[ht!]
    \centering
    \includegraphics[width=\textwidth]{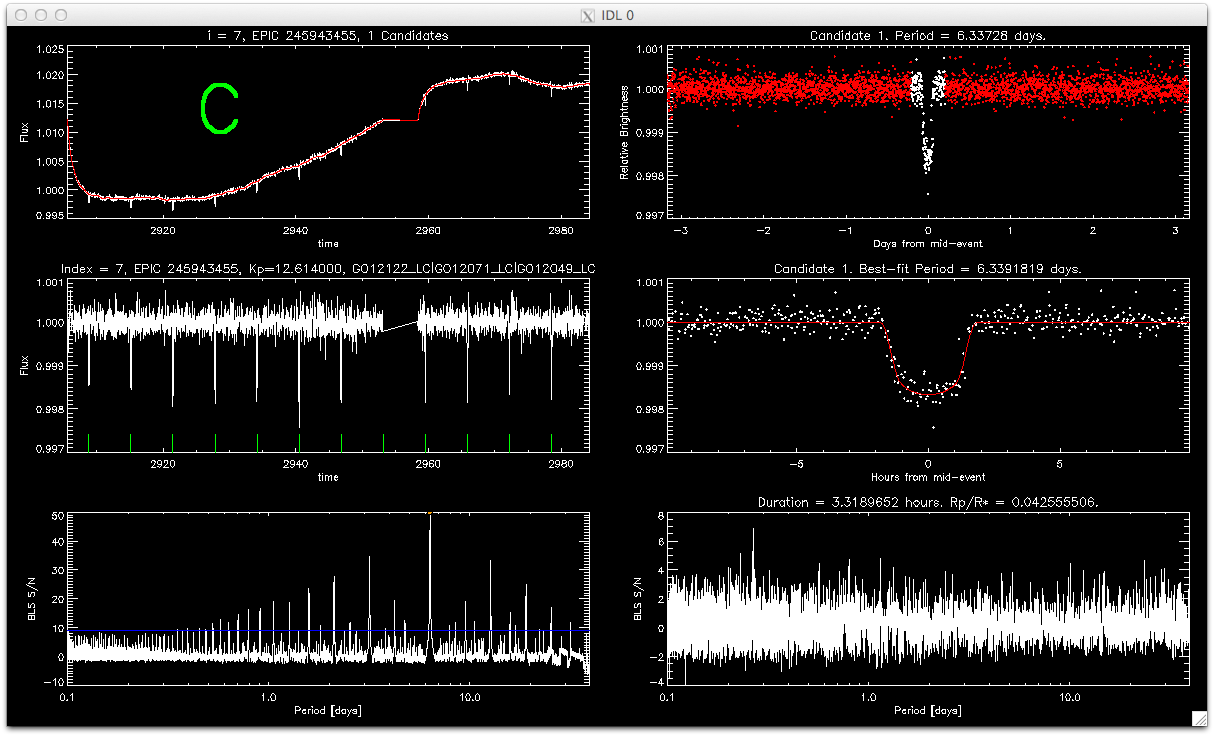}
    \caption{View of program used to initially classify targets in the triage stage (Section \ref{triage}). Top Left: unflattened light curve with classification label. Top Right: phase-folded light curve with transit event highlighted in white. Mid Left: Flattened light curve with transit events marked. Mid Right: a zoom-in on the transit event from the panel above. Bottom Left: BLS Periodigram of the signals found in the original light curve, shown in the top left panel. Bottom Right: BLS Periodigram after removal of the signal.}
    \label{fig:triage}
\end{figure*}

One of us (AV) performed triage by visually inspecting the light curve of each TCE and assigning it to one of three categories: ``C'' for planet candidate, ``E'' for eclipsing binary, or ``J'' for junk. Because of the large number of TCEs that must be triaged (several thousand typically per K2 campaign), triage was a rapid-fire process in which we separated those signals that do not remotely look like a planetary transit from those signals that could possibly be a candidate. At this stage, we were fairly liberal about which signals were labeled as candidates and were subsequently passed to more detailed vetting (described in Section~\ref{vetting}). The main goal of triage is to sort out all the obvious non-candidates - most TCEs are caused by instrumental artifacts, not astrophysical phenomena, so removing those TCEs massively reduces (more than 90\%) the number needed for further analysis. Figure~\ref{fig:triage} shows a screen capture of the program we used to perform triage. We used information from all six panels to determine whether the TCE could be caused by a planet.

\subsubsection{Vetting}\label{vetting}

\begin{figure*}[ht!]
    \centering
    \includegraphics[width=\textwidth]{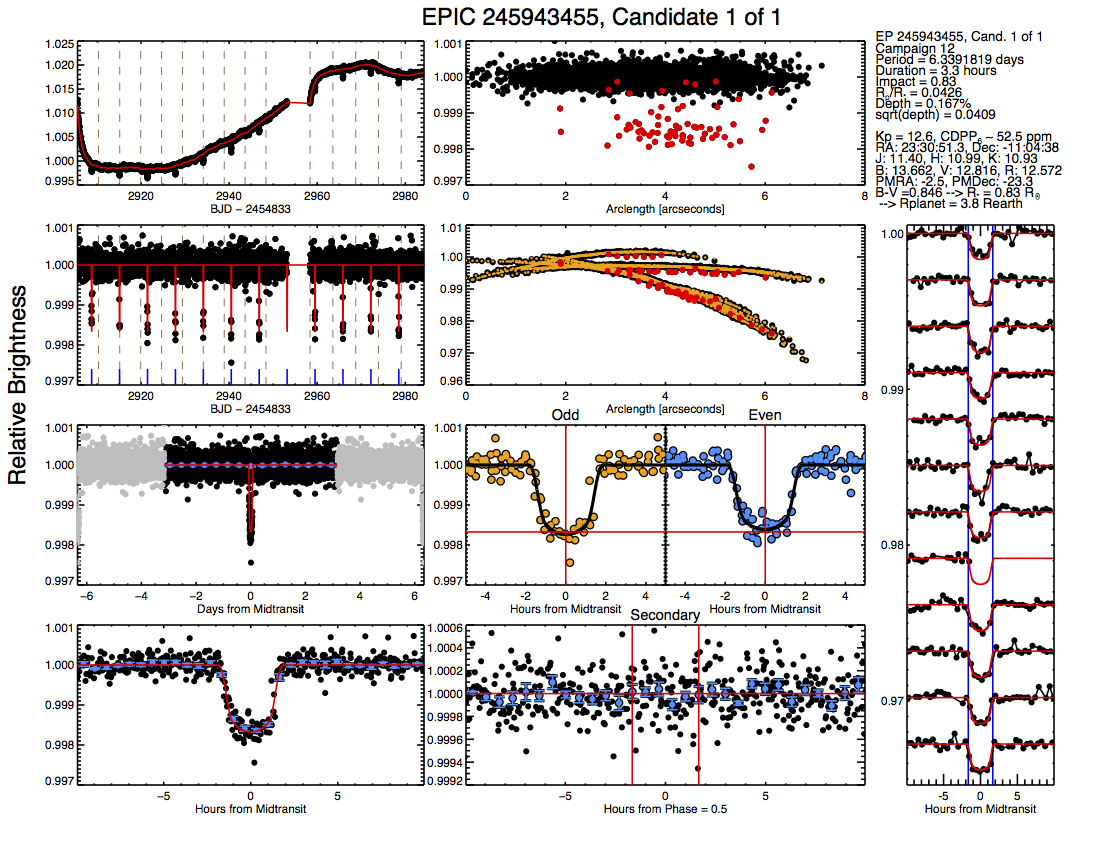}
    \caption{Sample diagnostic plots used in the vetting process (Section \ref{vetting}) for the planet candidate around EPIC 245943455. Left column, first and second rows: K2 light curves with and without low-frequency variability, respectively. The low-frequency variations are modeled in red in the first panel, while the best-fit transit model is shown in red in the second row. Vertical brown dotted lines denote the regions into which the light curve was separated to correct roll systematics. Left column, third and fourth rows: phase-folded, low-frequency corrected K2 light curves. The third row shows the full light curve with points more than one half-period from the transit shown in gray, while the fourth row shows only the light curve near transit. The blue points are binned data points and the red line is the best-fit transit model.  Middle column, first and second rows: arclength of centroid position of star versus relative brightness, after and before roll systematics correction, respectively. Red points denote in-transit data. In the second row, small orange points denote the roll systematics correction made to the data. Middle column, third row: odd transits are plotted in orange and even transits are plotted in blue. The black line is the best-fit model, the horizontal red line shows the modeled transit depth, and the vertical red line denotes mid-transit. This is useful for detecting binary stars with primary and secondary eclipses. Middle column, fourth row: light curve data in and around the expected secondary eclipse time (for zero eccentricity). The blue data points are binned data, the horizontal red line denotes a relative flux = 1, and the two vertical red lines denote the expected beginning and end of a secondary eclipse. Right column: Individual transits, vertically shifted from each other, with the best fit model shown in red. The blue lines denote the beginning and end of transit.}
    \label{fig:vetting1}
\end{figure*}

\begin{figure*}[ht!]
    \centering
    \includegraphics[width=\textwidth]{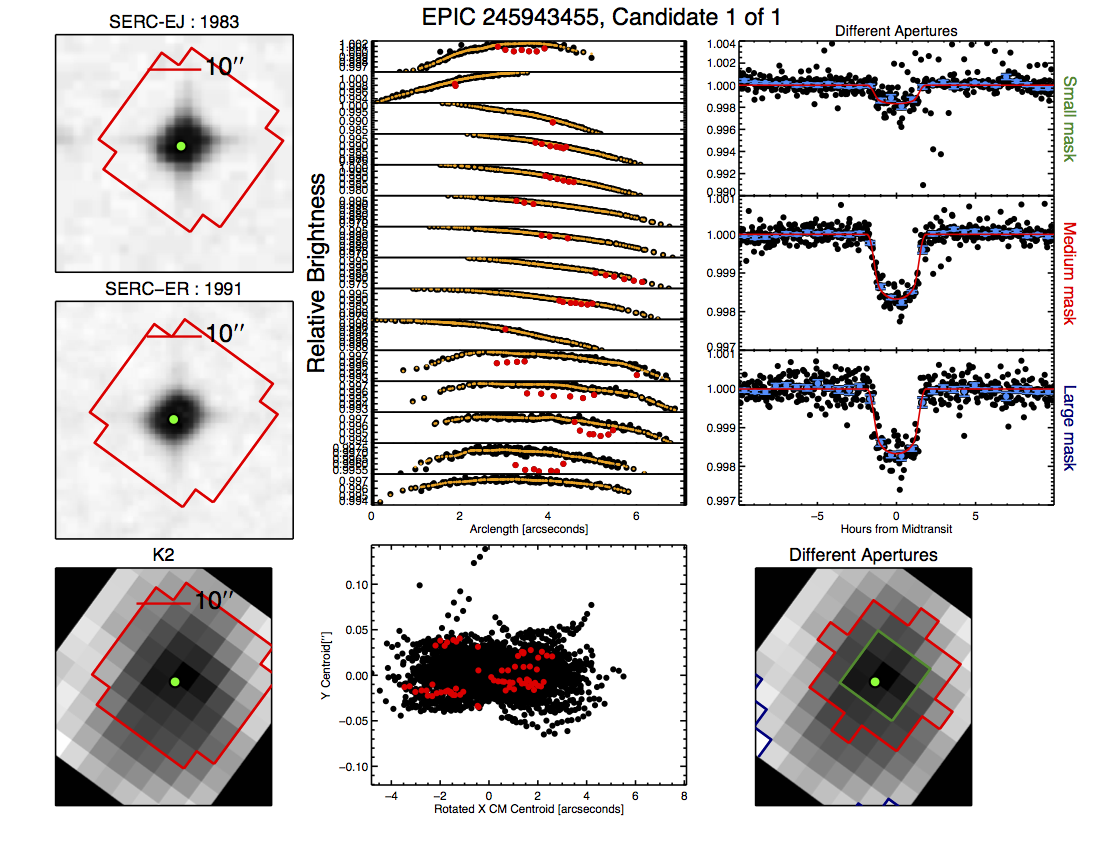}
    \caption{A second page of sample diagnostic plots used in the vetting process (Section \ref{vetting}) for the planet candidate around EPIC 245943455. Left column, first, second, and third rows: archival images from the Digital Sky Survey, the second Digital Sky Survey, and K2. Each has a scale bar at the top of the image and an identical red polygon that was the shape of the aperture chosen for reduction. The K2 image has been rotated to the same orientation as the Digital Sky Survey images.  Middle column, top row: panels of uncorrected brightness versus arclength, ordered chronologically and separated into the divisions in which the roll systematics correction was calculated. In-transit data points are shown in red and the orange points denote the brightness correction used to remove systematics. Middle column, bottom row: variations in the centroid position of the K2 image. In-transit points are shown in red. The discrepancy (in standard deviations) between the mean centroid position in transit and out-of-transit is shown on the right side of the plot. Right column, first row: the K2 light curve near transit as calculated using three different sized apertures: small mask (top panel), medium mask (middle panel), and large mask (bottom panel), each with the identical best-fit model in red and binned data points in blue. Aperture-size dependent discrepancies in depth could suggest background contamination from another star. Right column, third row: the K2 image with the three masks from the previous plot shown in green, red, and blue, respectively.}
    \label{fig:vetting2}
\end{figure*}

After we triaged the TCEs, we gave the signals we classified as planet candidates (those we labeled ``C'') more careful consideration. We used a large number of different diagnostics, which we show in Figures \ref{fig:vetting1} and \ref{fig:vetting2}. To ensure uniformity in the training set, we devised a set of rules to guide our determination of which signals were considered a candidate in our final training set. The rules were as follows:

\begin{itemize}
    \item A signal is a planet candidate until proven otherwise, either by astrophysical means (a strong secondary eclipse, odd/even transit depth differences, pixel-level diagnostics indicating the source of the transit is off-target, etc), the indication that the signal is due to instrumental systematic errors, or by violating one of the other rules.
    \item Any signal with a transit depth greater than 3\% is labeled an eclipsing binary. We identified many transit signals deeper than 3\% that are very likely eclipsing binary stars, even though we did not detect a secondary eclipse or odd/even depth differences. Since almost all known transiting planets have depths less than 3\%, we classified these signals as eclipsing binaries. We made an exception for Qatar-2 b, a known planet with an unusually deep transit \citep{bryan2012}.
    \item If the star was synchronously rotating with the orbit \textit{and} it has a v-shaped transit, it is labeled an eclipsing binary.
    \item Any TCEs with phase modulations that we deem to be caused by tidally distorted stars, relativistic beaming, or other astrophysical phenomena are labeled eclipsing binaries.
    \item Any TCEs that are so ambiguous that we cannot decide whether they are viable planet candidates or false positives are removed from the training set. 
    \item TCEs caused by single transits are removed from the training set.
    \item Disintegrating planets, in particular WD 1145+017 b \citep{wd1145} and K2-22 \citep{k2-22}, are removed from the training set. 
    
\end{itemize}
We also removed all TCEs from several K2 campaigns from our training set entirely. We removed Campaign 0 and Campaign 11 because they were pointed towards crowded regions of the sky (with a large number of ambiguous but likely false positive signals). Campaign 0 was also unusual among K2 campaigns given its shorter duration. We also ignored Campaign 9, which was pointed at the galactic bulge and focused on microlensing. 
 
 {\ron At this stage, we also corrected by hand some cases where our pipeline mis-identified a TCE's orbital period (usually at half or double the true value).} 


\subsection{Preparing Input Representations}\label{preprocessing}

Once each TCE had been identified and labeled, we processed the TCE into a standardized input representation to be sent into the neural network. Much of this section follows the strategy laid out by \citet{shallue}. We started by removing long-term variability from each light curve. Although we have previously removed short timescale instrumental systematics from the light curve \citep[following][]{vj14}, the light curves still have long-term signals caused either by stellar variability or long-term instrumental systematics that must be removed. 

Following \citet{shallue}, we fit a spline to the light curve to remove low-frequency variability from the star. 
The data were then phase-folded along the TCE's period such that each transit event lined up and was centered. {\ron To remove any unusual data points, such as those caused by cosmic rays or those taken when solar system planets or large asteroids passed near the target star, upward outliers three standard deviations or more from the mean (computed robustly) of the light curve were removed.} 

We then converted the light curve into a standardized input representation for the neural network.
Like \citet{shallue}, we binned the data in two ``views'': a global view, which shows the characteristics of the light curve over an entire orbital period, and a local view, which shows the shape of the transit in detail. The global view is the entire phase-folded light curve with each data point grouped into one of 701 bins and took the median of all data points within each bin. The local view is a close-up of the transit event, spanning four transit durations on either side of the event. We grouped the data points from the phase folded light curve within this time range into one of 51 bins and took the median within each bin. Many light curves had gaps in the data that caused some bins to have zero data points. In those cases, we determined values for the bins without any data points by linearly interpolating between nearby bins with data points. To normalize the light curves, we re-scaled the binned values such that the median point was set to zero and the minimum point (usually the deepest point in transit) was set to -1. We deviated from \citet{shallue} by using smaller numbers of bins for both the local and global views, and also by interpolating the data. These modifications were necessary for K2 data because there are fewer data points in K2 light curves than in \Kepler, making it more likely for some bins to have no data points.

\begin{figure*}
    \centering
    \includegraphics[width=6.5in]{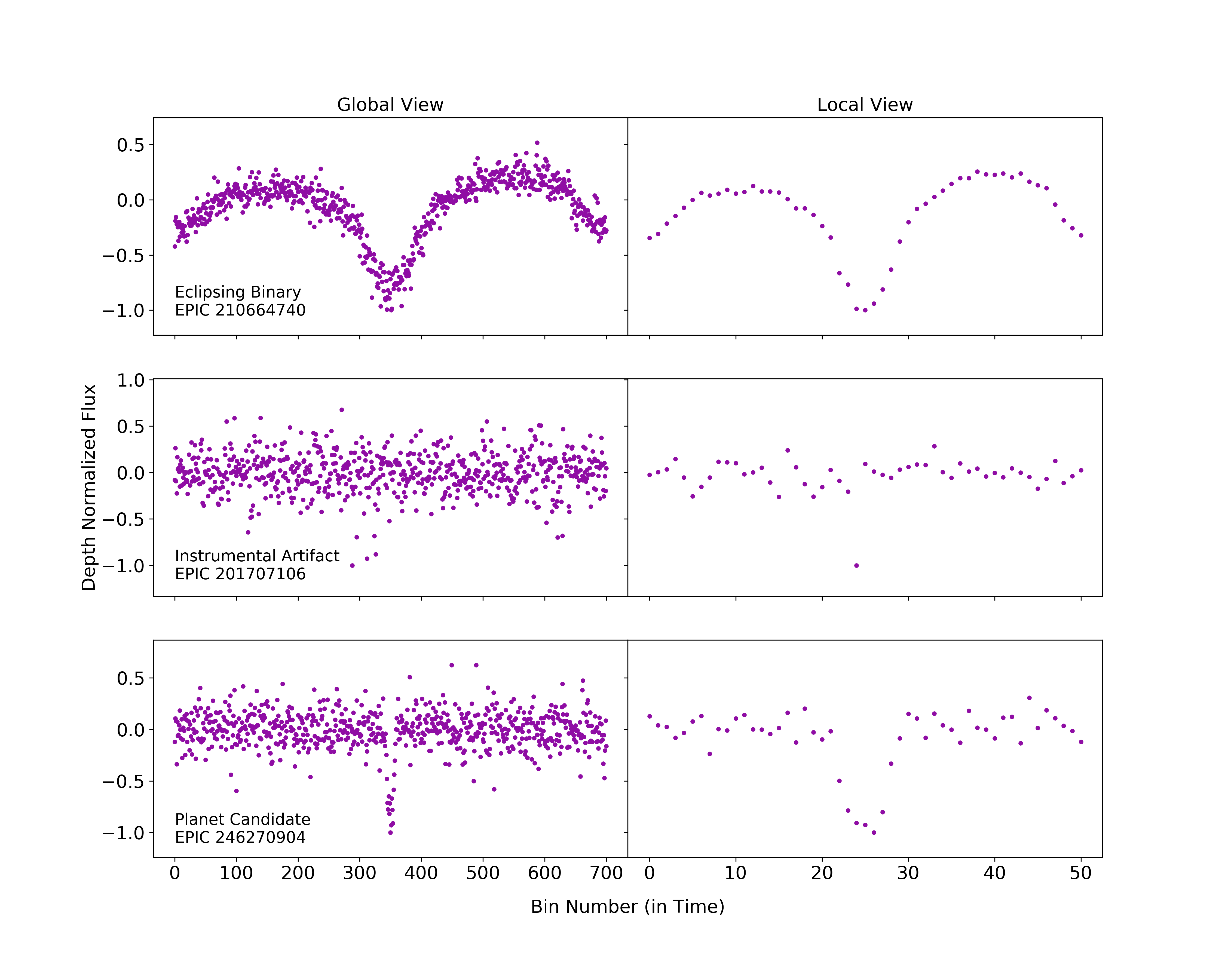}
    \caption{Each TCE was represented to the model as a phase-folded, depth-normalized light curve of equal length. Each TCE had two views: a ``Global View'' to represent the entire period and a ``Local View'' showing a close-up of the possible transit. These are shown in left and right columns, respectively, for an example of each category of TCE.}
    \label{fig:glob_loc}
\end{figure*}

Figure~\ref{fig:glob_loc} shows examples of the global and local views with different classes of signals. The top row shows an eclipsing binary, the middle row shows an instrumental artifact, and the bottom row shows a planet candidate. Each example has a possible transit event that is more obvious in the local view.

We also input several scalar features to our neural network, in particular, the best-fit values of the planet/star radius ratio ($R_p/R_\star$) and the transit impact parameter from a least squares fit to the phase-folded TCE light curve using a \citet{mandelagol} model. These two features allow the neural network to use information about the depth of the transit (which is a non-trivial function of the impact parameter and planet/star radius ratio, see \citealt{seager2003}).
Including depth as a scalar feature rather than leaving the depth of the light curve un-normalized helps the machine learning algorithm learn more efficiently by decoupling the transit shape and depth. Including these scalar features is another difference between our method and that of \citet{shallue}

{\ron The total dataset used for training comprised of 27,634 TCEs. We randomly shuffled and divided the data into three subsets: training (80\%, 22,105 TCEs), validation (10\%, 2,774 TCEs), and test (10\%, 2,755 TCEs).} We used the test set to evaluate final model performance and we used the validation set to check performance to optimize the metaparameters\footnote{We use the term \textit{metaparameters} for choices that modify the model architecture (e.g. number of layers) or training algorithm (e.g. learning rate).}. The random shuffling allows data from each campaign to be spread evenly among the subsets, and the separate test set allows a cleaner final performance evaluation since it is not used to train the model or select the metaparameters.

\section{Neural Network Model}\label{nn model}

\subsection{Architecture}\label{architecture}

We based our neural network architecture on the \texttt{AstroNet} model \citep{shallue}, which is implemented in TensorFlow, an open source machine learning framework \citep{abadi2016tensorflow}. Our code is available online.\footnote{\url{https://github.com/aedattilo/models_K2}}

\begin{figure}
    \centering
    \includegraphics[width=0.5 \textwidth]{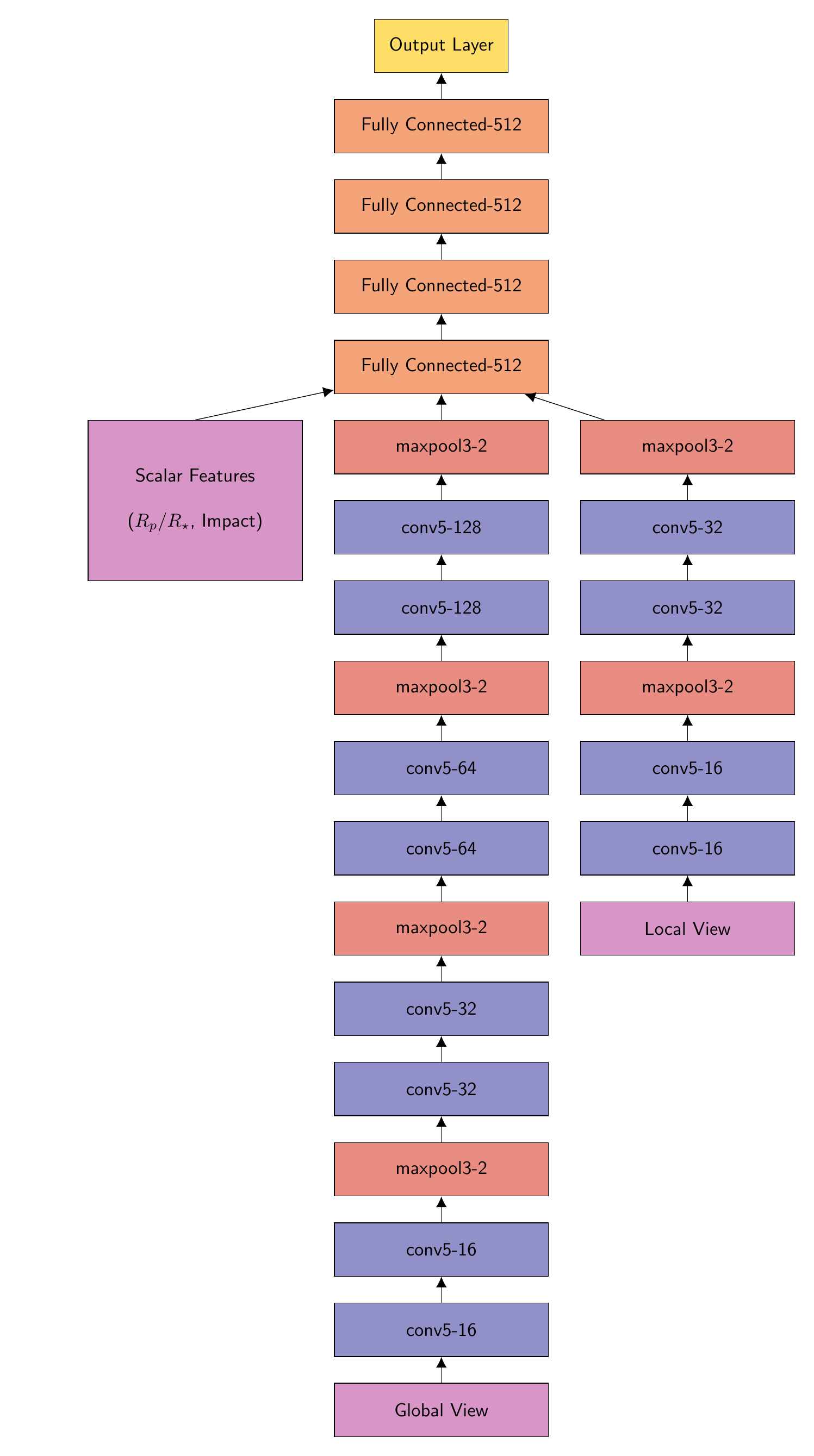}
    \caption{The architecture of our best performing neural network model. Convolutional layers are denoted conv\textlangle kernel size\textrangle -\textlangle number of feature maps\textrangle , max pooling layers are denoted maxpool\textlangle window length\textrangle -\textlangle stride length\textrangle , and fully connected layers are denoted FC-\textlangle number of units\textrangle .}
    \label{fig:arch_diagram}
\end{figure}

We started by attempting to train the original \texttt{AstroNet} architecture on our new K2 input data and training set labels. This initial attempt to train the neural network {\ron never converged}. We therefore optimized the neural network architecture and training parameters by training on a data set we already knew worked: the \Kepler\ data set and the Autovetter training labels used by \citet{shallue}. We truncated \Kepler\ light curves to 80 day segments to mimic the shorter time span of K2 observations of a given target star. We only included targets with orbital periods less than 80 days, again in order to mimic the K2 set of planet candidates. This \Kepler-lite model ran best with a learning rate of $\alpha$ = $10^{-4}$ and the architecture described in Figure~\ref{fig:arch_diagram} (excluding the two scalar features). We then used this successful architecture and learning rate to train with the K2 data set and training examples.

Figure~\ref{fig:arch_diagram} represents the architecture of our final K2 model, called \texttt{AstroNet-K2}, which is a one-dimensional convolutional neural network with max pooling. Both global and local views are used, as described in Section~\ref{preprocessing}, and are separately passed through convolutional columns before combining them in shared fully connected layers. The output layer used a sigmoid function whose range is (0,1). The output of the model is a prediction of the probability that the TCE is a planet candidate. Output values close to 1 signify high confidence that the TCE is a planet candidate while output values close to 0 signify high confidence that the TCE is a false positive. As discussed previously, we also included $R_{p}/R_{\star}$ and impact parameter, which we found to help the model.


\subsection{Training}\label{training}
{\ron We trained the model using the 22,105 TCE light curves in the training set (80\% of all the labeled examples) for 4,000 steps.} We used the Adam optimization algorithm \citep{kingma2014adam} to minimize the cross entropy error function over the training set. We used a learning rate of $\alpha = 10^{-4}$ and a batch size of 64. We augmented the training data by randomly {\ron reversing the time order of light curves during the training process. Since true planet transits are symmetric in time, reversing the time order produces a new light curve with the same label as the original light curve. Therefore, this operation effectively doubled the size of our training set.}

\section{Evaluation of the Neural Network Performance}\label{eval}
We evaluated the performance of our neural network with respect to several metrics. To ensure any small variations between models were minimized, we evaluated the performance using an averaged model. We trained ten separate models with the same parameters and averaged their final predictions for each TCE. {\ron We computed all metrics in this section over the test set (rather than the training or validation sets) to avoid using any data} that was used to optimize the model or the metaparameters.

\subsection{Metrics/Histogram}

\begin{figure}
    \centering
    \includegraphics[width=0.48\textwidth]{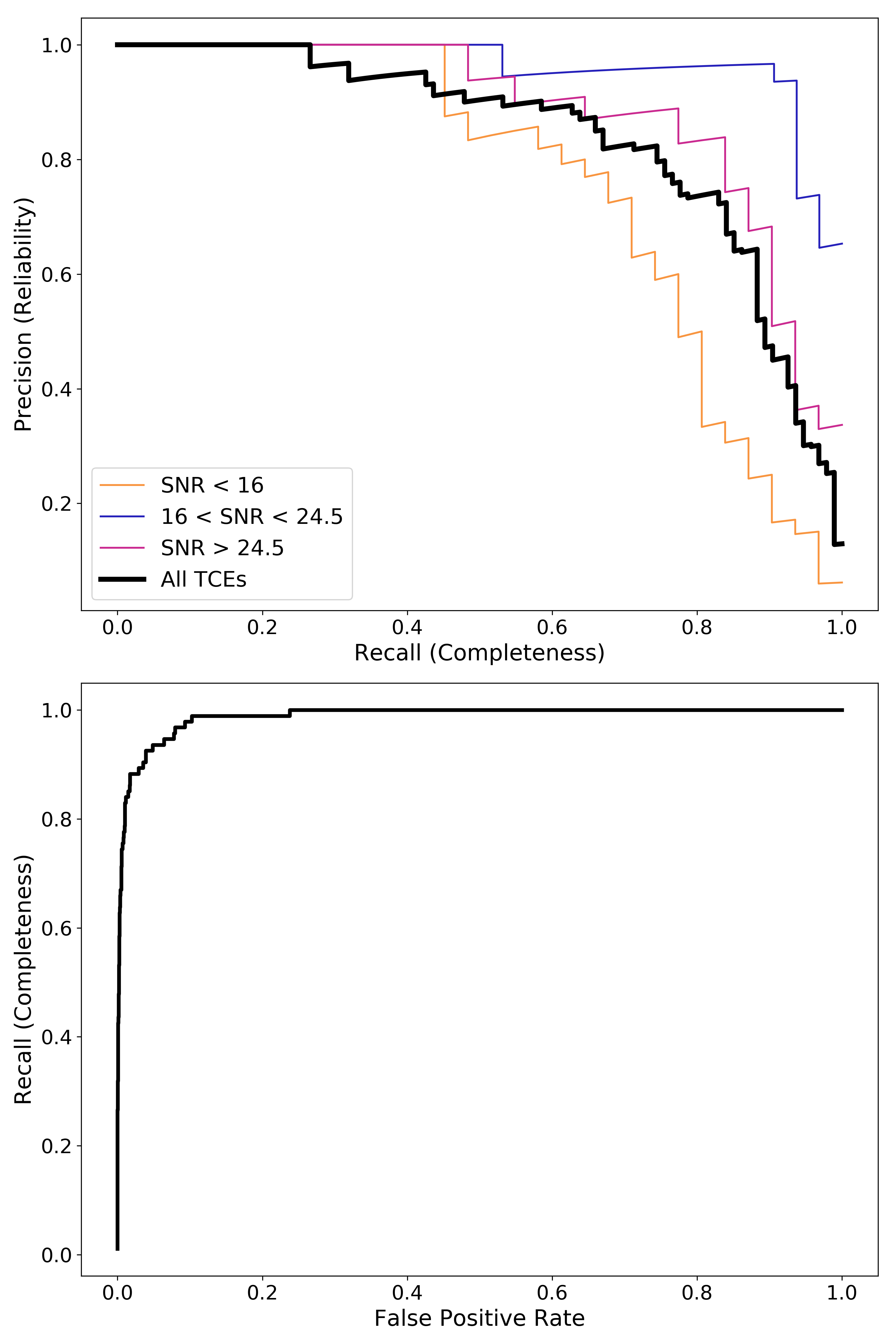}
    \caption{The top graph shows the fraction of planets that the model classified as planets (recall) versus the fraction of correctly classified planets (precision). This curve shows the trade-off between having no false positives (high precision) and identifying all planet candidates (high recall). In order for the model to recall most or all of the planet candidates, there will be a large pool of false positives contaminating the planet candidate sample. {\ron We split the test set TCEs into three different groups based on SNR such that each group had the same number of planet candidates. Our model performed better on TCEs in the two highest SNR bins ($\mathrm{SNR}>16$) compared to the full test set.} The bottom graph, also known as the receiver-operator characteristic (ROC) curve, is {\ron the} curve from which the AUC value is calculated. The ROC curve combines information about the recall (true positive rate) of the model with information about the ability to recognize false positives (the false positive rate). Our model is highly successful at identifying false positives, so the AUC value is high.}
    \label{fig:metrics}
\end{figure}

There are many ways to evaluate the performance of a classifier, so we use a few different metrics to describe the performance of our model. The metrics we included for evaluation were:
\begin{itemize}
    \item Accuracy: the fraction of signals the model classified correctly.
    \item {\ron Precision (reliability):} the fraction of all classified planet candidates that were true positives (labeled planet candidates).
    \item {\ron Recall (completeness):} the fraction of total labeled planet candidates that the model classified as planet candidates.
    \item {\ron False positive rate: the fraction of total labeled false positives that the model classified as planet candidates.}
    \item AUC (area under the receiver-operator characteristic curve; {\ron see Figure~\ref{fig:metrics}, bottom panel}): the probability that a randomly selected planet candidate achieves a higher prediction value than a randomly selected false positive.
\end{itemize}
{\ron The values of the first four metrics depend on the classification threshold we choose for our model. If we consider predictions greater than 0.5 to be classified as planet candidates and predictions less than 0.5 to be classified as false positives, then our averaged model attains an accuracy of 97.84\%. The value of AUC is independent of the choice of classification threshold, and our averaged model attains an AUC of 98.83\%.}

{\ron Figure~\ref{fig:metrics} shows the precision vs. recall curves for our averaged model. Precision vs. recall curves show our model's trade-off between sensitivity and specificity: if we tune the classification threshold to have few false negatives, then we will still have many false positives. If we take the precision to be 1, i.e. no false positives, the recall is only 0.2, which means we are missing 80\% of the planet candidates.} This is expected because of the low rate of planet candidates in the data set and because occasionally, systematic noise can closely mimic planet signals. The model is able to classify the true planet candidates well, but because there is a large pool of false positives, even relatively high accuracy rates can lead to significant contamination. 

We also investigated how our model performs when classifying TCEs at different Signal-to-Noise Ratios (SNR)\footnote{We use the significance of the TCE's BLS detection (as defined by \citealt{v16}) as the SNR of the candidate.}.  In addition to calculating precision vs. recall curves over the full test set, we also calculated these curves for several sub-groups of the test set divided in SNR. We found that our model performs worst on low SNR TCEs ($\mathrm{SNR}<16$), as the area under the precision vs. recall curve on is lowest for this subgroup. Somewhat surprisingly, we found that our model performs best on signals in the middle SNR bin, with $16<\mathrm{SNR}<24.5$. We suspect this is because the highest SNR TCEs include a larger fraction of eclipsing binary false positives which the model might misclassify, reducing its overall accuracy in that group.

\begin{figure*}[ht!]
    \centering
    \includegraphics[width=7in]{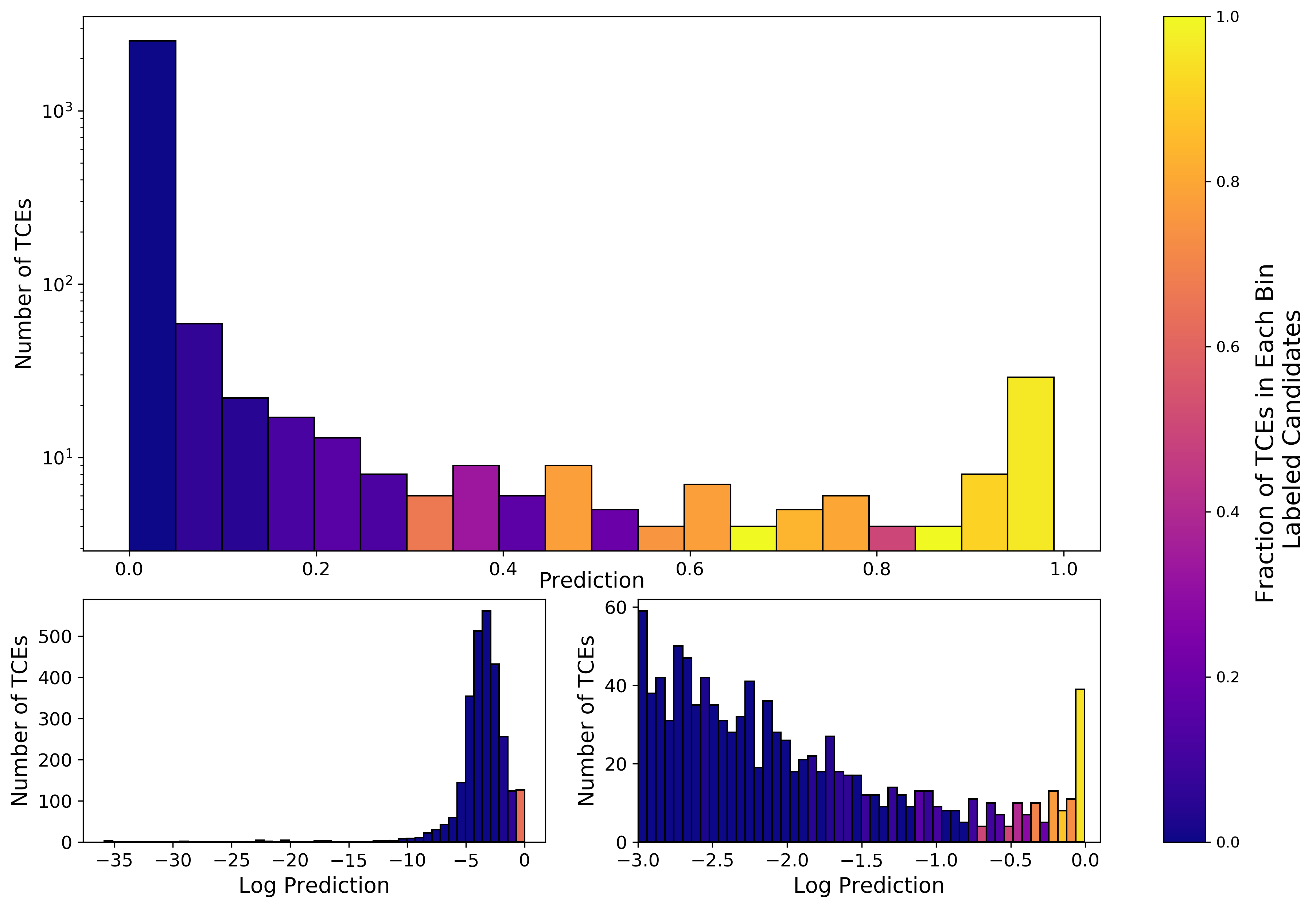}
    \caption{Histogram of test data predictions. The color of each bin represents the fraction of TCEs that were labeled as candidates within that prediction range. Top: All data within the test set, represented in a log y scale. Note that the bin at 0.8 only contains {\ron four TCEs, and the model predicted two non-planet candidates and two planet candidates in that bin.} Bottom left: Histogram of all of the data within the test set, presented by the logarithm of their prediction. Bottom right: Same as bottom left, but zoomed in to view only TCEs with predictions greater than $10^{-3}$, shown with log scale of predictions. Most data still have very low predictions.}
    \label{fig:hist}
\end{figure*}

How well the model classified signals can also be visualized by the predictions the model gave to the signals in the test set. Figure~\ref{fig:hist} shows a histogram of predictions given to the test set. A prediction of 0 represents the neural network predicting that the TCE was not a planet, while a prediction of 1 represents the neural network predicting the TCE was a planet. {\ron Ignoring the color, the top panel shows that a majority ($\sim$88\%) of the TCEs received a prediction of 2\% or below.} The color then represents a ``truth value'' to the neural network's predictions: within each bin, the color represents how many TCEs were planet candidates. A yellow bin shows that all of the TCEs that received predictions within that range were planet candidates while a blue bin shows that all the TCEs in that bin were not planet candidates. This histogram demonstrates that the neural network works as expected: on the right hand side, the colors are yellow and a majority of the TCEs that received high predictions were planet candidates; in the middle there are a mix of colors, where the TCEs were a mixture of planet candidates and other labels; and on the left hand side the bins are blue and a majority of the TCEs were not planet candidates.

The bottom two panels show the same histogram in log scale. The left bottom panel shows that the neural network was very good at assigning non-planet TCEs very low predictions, while the right bottom panel shows the distribution of TCEs with a cut of $10^{-3}$.

\subsection{How can we use this today?}
Our model is not yet ready to completely automatically detect and identify planet candidates: when it is run on a group of TCEs, it incorrectly identifies too many false positive signals as planet candidates, which contaminates the planet candidate sample. If we tune our threshold for identifying planet candidates so that we successfully recall more than 90\% of the planet candidates in our test set, the resulting planet candidate sample has a precision of only 40\%, meaning that true planet candidates are outnumbered by false positives. %

However, even though our model is not ready to classify planet candidates and false positives without human intervention, it can be powerful when combined with some human oversight. For example, we could cull out obvious false positives in a set of TCEs by performing a conservative cut on the prediction value output by our model. If we choose to discard all TCEs for which our model assigned a prediction less than 0.01, we would discard only one real planet in our test set, but cut out 85\% of the false positive TCEs. This is a straightforward way of decreasing a human's workload by a factor of 6 when identifying planet candidates from a list of TCEs. 


In this way, we can use our neural network like a sieve to weed out the least planet-like signals, allowing humans to spend more time scrutinizing strong candidate planets, not the obvious false positives. This is particularly useful for identifying new planets among a large number of unclassified TCEs by shrinking the pool of TCEs a human astronomer would have to look at and dramatically speeding up the amount of time it takes to identify new planets. 

\section{Testing on new TCEs}\label{new}



Because the model is successful at culling large amounts of un-labeled TCEs, we ran it on a new set of TCEs from K2. Recall from Section~\ref{search} that the data used for the training set only included the first TCE detected around each star. {\ron These new TCEs were the other signals found when the BLS pipeline iteratively removed the strongest signals from each target.} We focused on the set of un-classified TCEs that were from targets where the first signal was labeled ``J'' or ``E.''  This resulted in a subset of 22,050 TCEs to search through.



The model ranked 826 of these TCEs above 1\%. We noticed that many (229) of these new candidates were from Campaign 12, likely because Mars passed through \Kepler's field of view during the campaign, causing additional systematic effects in the light curves. These additional systematic effects often were the first signals identified by our transit search, hiding any real transit events. We decided to focus our attention on these new, high-prediction value TCEs from Campaign 12, which we list in Table~\ref{tab:c12}. We chose two of these signals, towards \thisstarone\ and \thisstartwo, for follow-up observations and validation because of their high prediction scores and bright apparent magnitudes.

\begin{deluxetable*}{ccccccccccc}
\tablewidth{0pt}
\tablecaption{\label{ta} New Highly Ranked TCEs from Campaign 12}
\tablehead{
  \colhead{EPIC} &
  \colhead{TCE} &
  \colhead{Period} &
  \colhead{T0} &
  \colhead{Duration} &
  \colhead{$(R_p/R_\star)^2$} &
  \colhead{Impact} &
  \colhead{Kepler} &
  \colhead{SNR} &
  \colhead{Prediction} &
  \colhead{Notes} \\
  \colhead{ID} &
  \colhead{Number} &
  \colhead{(days)} &
  \colhead{(BJD-2454833)} &
  \colhead{(hours)} &
  \colhead{} &
  \colhead{Parameter} &
  \colhead{Magnitude} &
  \colhead{} &
  \colhead{} &
  \colhead{}
}
\startdata 
    246222846 & 3 & 1.962674 & 2907.283 & 0.8855 & 0.0226 & \ron{0.5094} & 9.78 & 11.68 & 0.3197 & $a,b$  \\ 
    246078672 & 4 & 2.503312 & 2907.016 & 1.9320 & 0.0119 & 0.2679 & 12.39 & 9.04 & 0.3534 & $e,c,g$ \\ 
    246047423 & 2 & 9.775235 & 2915.073 & 3.9934 & 0.0346 & 0.9430 & 13.55 & 10.47 & 0.4396 & $a$ \\
    246354333 & 2 & 9.753945 & 2915.117 & 4.5894 & 0.0223 & 0.9651 & 14.13 & 10.55 & 0.4675 & $a$\\
    246082797 & 2 & 26.348500 & 2930.717 & 6.2716 & 0.0521 & 0.9438 & 13.71 & 9.58 & 0.5255 & $a$ \\
    246006388 & 2 & 14.451154 & 2905.740 & 4.0048 & 0.0322 & 0.0016 & 13.78 & 10.59 & 0.5720 & $a$ \\
    246219137 & 2 & 28.869769 & 2905.745 & 2.9950 & 0.0403 & \ron{0.7035} & 13.45 & 9.44 & 0.6172 & $a$ \\
    246244554 & 2 & 15.810046 & 2908.859 & 3.4777 & 0.0241 & 0.2466 & 13.36 & 9.21 & 0.6263 & $a$ \\
    246414947 & 2 & 0.359275 & 2905.721 & 0.8636 & 0.0828 & \ron{0.0005} & 18.71 & 23.11 & 0.6913 & $d$  \\
    246163416 & 5 & 0.876857 & 2905.853 & 1.1297 & 0.0514 & 1.01823 & 13.48 & 16.61 & 0.8281 & $e,g,j$ \\
    246260670 & 6 & 3.853670 & 2907.801 & 2.2740 & 0.0823 & 0.8346 & 15.28 & 52.32 & 0.8360 & $e,h,k$ \\
    246140166 & 2 & 2.741774 & 2907.923 & 1.5104 & 0.0412 & 0.6729 & 15.92 & 9.92 & 0.9291 & $e,g,k$ \\ 
    245944045 & 2 & 4.203831 & 2908.692 & 2.0121 & 0.0237 & 0.0183 & 13.56 & 13.17 & 0.9474 & $e,g,i/k$ \\ 
    246151543 & 6 & 13.123778 & 2916.131 & 3.6658 & 0.0205 & 0.0364 & 13.17 & 13.79 & 0.9839 & $e,f,g,i$ \\ 
\enddata
\tablenotetext{$a$}{Likely instrumental systematic.}
\tablenotetext{$b$}{TCE host star saturated on the \Kepler\ detector, introducing additional systematics.}
\tablenotetext{$c$}{Validated as K2-294 b}
\tablenotetext{$d$}{Likely astrophysical signal, but transit shape is reminiscent of an astrophysical false positive.}
\tablenotetext{$e$}{Strong planet candidate.}
\tablenotetext{$f$}{Validated as K2-293 b}
\tablenotetext{$g$}{Possible super-Earth/sub-Neptune sized planet.}
\tablenotetext{$h$}{Possible sub-Saturn sized planet.}
\tablenotetext{$i$}{G-dwarf host star.}
\tablenotetext{$j$}{M-dwarf host star.}
\tablenotetext{$k$}{K-dwarf host star.}

\label{tab:c12}
\end{deluxetable*}

\subsection{Observations}\label{observations}

\subsubsection{K2 Light Curve}\label{lightcurve}


\thisstarone\ and \thisstartwo\ were both observed by \Kepler\ during Campaign 12 of the K2 mission. Campaign 12 observations took place between 2016 December 15 and 2017 March 04, and were pointed towards an equatorial field in the constellation Aquarius near the southern galactic cap. Peculiarities of these particular K2 observations included a safe mode event that took place during the middle of the campaign, which led to the loss of 5.3 days of data, and the fact that the solar system planet Mars passed through the field of view, scattering light across the focal plane, and contaminating some targets strongly with either reflected light or saturated bleed trails{\ron \footnote{\url{https://archive.stsci.edu/k2/data_release.html}}}. After the data were downlinked from the telescope and processed by the \Kepler\ pipeline, we downloaded the data, produced light curves, removed systematics, and searched for transits as described in Section \ref{training set}. 

Parts of the light curves for \thisstarone\ and \thisstartwo\ were both significantly affected by the passage of Mars through \Kepler's field of view. For a brief period of time for both targets, the typical background flux in pixels near the target stars increased by a factor of about 150. This spike in background flux from Mars was not properly handled by the \Kepler\ pipeline's background subtraction algorithms, and the effects of this propagated forward through our pipeline, causing a poor systematics correction near the time of the contamination. As a result, while the majorities of the light curves of \thisstarone\ and \thisstartwo\ had high-quality systematics corrections, some significant systematics remained after our routines. 

{\ron As described in Section~\ref{search}, we} then passed these systematics-contaminated light curves to our our BLS pipeline to search for transits. Our pipeline, which searches for the most significant drops in flux in the light curve, first identified spurious transit signals related to the passage of Mars through the field of view. Our pipeline then proceeded to iteratively remove the strongest signals in the light curve and re-search for transits, eventually identifying the two likely transiting planet signals in the light curves. However, during triage, because we only looked at the strongest signal identified by our pipeline for any given star, we did not notice the two transit signals buried among the systematics in the light curves of \thisstarone\ and \thisstartwo\ until our neural network identified the signals as likely planet candidates {\ron when the new TCEs were passed through the trained neural network}. We inspected the signals by eye, and performed standard pixel-level diagnostics \citep{v16, mayo}, and found that the signals were consistent with genuine transiting planets. 

Once we identified these signals, we produced improved light curves from the K2 data. First, we clipped out the segments of the light curve that were strongly affected by the passage of Mars through the field of view. We then re-derived the systematics correction by simultaneously fitting for the transit shape, long-term stellar variability, and K2 roll-dependent systematic effects \citep[following][]{v16}. We used these improved light curves for the rest of the analysis on these two planet candidate systems. 

\subsubsection{Spectroscopy}\label{spectroscopy}

We observed \thisstarone\ and \thisstartwo\ with the Tillinghast Reflector Echelle Spectograph (TRES) on the 1.5 meter telescope at Fred L. Whipple Observatory on Mt. Hopkins, AZ. Our observations were obtained with resolving power $\lambda/\Delta\lambda$ = 44,000, and moderate signal-to-noise (which varied between 25 to 35 per resolution element). We obtained one observation of \thisstarone\ and two observations of \thisstartwo. The two observations of \thisstartwo\ were taken at different orbital phases, {\ron 0.26 and 0.74}, but showed no evidence for large radial velocity variations, ruling out eclipsing binary stars as a possible explanation for the transit signal we see. We determined spectroscopic parameters for these two stars using the Stellar Parameter Classification (SPC) code \citep{buchhave, buchhave14}, and report the results in Table \ref{bigtable}. 

\subsubsection{High Resolution Imaging} \label{speckle}

We also observed \thisstarone\ and \thisstartwo\ with the \Alopeke\ speckle imager on the Gemini-N telescope on Maunakea, HI. \Alopeke\ works by taking many images of a target star with very fine spatial sampling with exposure times fast enough (60 ms) to ``freeze'' atmospheric turbulence. We observed each star in two different narrow bandpasses: one centered around 562 nm and one centered around 832 nm. We processed the data by reconstructing the image in Fourier space, a standard method for speckle image processing \citep{howell2011}. We find no evidence for nearby stars to either \thisstarone\ or \thisstartwo. In the blue bandpass, the resolution is 17 mas, at which separation we rule out roughly equal brightness companions. At greater separations of about 0\farcs05 arcseconds, we can rule out companions at deeper contrasts of 4-6 magnitudes in the redder bandpass. 

We also searched for evidence of nearby stars using \textit{Gaia} Data Release 2 \citep{gaiamission,gaiadr2}. We queried all entries in the \textit{Gaia} catalog within 5 arcseconds of the two stars, and found no additional sources nearby. We also checked the significance of excess noise in the astrometric solution for this star, which in some cases can indicate the presence of an unresolved companion \citep{rizzuto18}. Neither star showed any evidence for excess astrometric noise, supporting the idea that these are both single stars and limiting false positive scenarios for their transiting planet candidates.  

\subsection{Analysis}
\subsubsection{Stellar Parameters}

We estimated the fundamental stellar parameters for \thisstarone\ and \thisstartwo\ using the \texttt{isochrones} package \citep{isochrones}, which performs a fit of input constraints (including spectroscopic paramters, parallax, apparent magnitudes, etc) to stellar evolutionary models. We included our measurements of the stellar surface gravity $\log{g_{\rm cgs}}$, metallicity [M/H], and effective temperature $T_{\rm eff}$ from our SPC analysis of the TRES spectra, measured trigonometric parallaxes of the two stars from \textit{Gaia} DR2, and $J$, $H$, and $K$ band apparent magnitudes from the Two Micron All Sky Survey (2MASS, \citealt{twomass}). The results are reported in Table \ref{bigtable}. 

\subsubsection{Transit Fitting}

We next estimated the transit parameters for \thisplanetone\ and \thisplanettwo. Our analysis here closely follows that of \citet{mayo}, which we describe in brief. In particular, for each system, we flattened the K2 light curve by dividing away the best-fit stellar variability model from our simultaneous fit. We then fitted the resulting, high-pass-filtered light curve with a transit model (as implemented by the \texttt{batman} package, \citealt{batman}), using quadratic limb darkening coefficients parameterized as suggested by \citet{kippingld}. We used the affine invariant ensemble sampler \texttt{emcee} \citep{emcee} to explore parameter space and determine the best-fit transit parameters and their uncertainties. In the case of \thisstartwo, we imposed a Gaussian prior on the stellar density determined by our analysis using \texttt{isochrones}, which combined with Kepler's third law and an assumption that \thisplanettwo's orbit is circular (as expected from tidal circularization) helps constrain the transit parameters. The results of the fits are reported in Table \ref{bigtable}

\begin{deluxetable*}{lcc}
\tablecaption{Parameters for \thisplanetone\ and \thisplanettwo \label{bigtable}}
\tablewidth{0pt}
\tablehead{
  \colhead{Parameter} & \colhead{\thisplanetone}  & \colhead{\thisplanettwo} \\
  \colhead{} & \colhead{K2-293 b} & \colhead{K2-294 b} }
\startdata
\emph{Stellar Parameters} & \\
Right Ascension & 23:26:25.52 & 23:28:12.4  \\
Declination & -06:01:40.35 & -6.027876 \\

$M_\star$~[$M_\odot$] & $0.958^{+0.036}_{-0.034}$ & $0.987^{+0.034}_{-0.030}$ \\
$R_\star$~[$R_\odot$] & $0.947^{+0.089}_{-0.055}$ & $1.20^{+0.13}_{-0.12}$ \\
Limb darkening $q_1$~ & $0.39^{+0.34}_{-0.26}$ & $0.58^{+0.29}_{-0.32}$ \\
Limb darkening $q_2$~ & $0.39^{+0.37}_{-0.27}$ & $0.52^{+0.32}_{-0.34}$ \\

$\log g_\star$~[cgs] & 4.52 $\pm$ 0.13 & 4.20  $\pm$ 0.10 \\
\meh & 0.218 $\pm$ 0.080 & 0.194 $\pm$0.080\\
$T_{\rm eff}$ [K] & 5532 $\pm$ 78& 5612 $\pm$ 50 \\
 & & \\
 
\emph{New Planet Parameters} & \\
Orbital Period, $P$~[days] & $13.1225^{+0.0011}_{-0.0012}$ & $2.50387^{+0.00022}_{-0.00023}$ \\
Radius Ratio, $(R_P/R_\star)$ & $0.0228^{+0.0020}_{-0.0012}$ & $0.01262^{+0.00076}_{-0.00081}$ \\
Scaled semi-major axis, $a/R_\star$ & $22.9^{+3.0}_{-7.7}$ & $6.49^{+0.50}_{-0.65}$ \\
Orbital inclination, $i$~[deg] & $88.8^{+0.8}_{-1.9}$ & $85.5^{+2.1}_{-2.0}$ \\
Transit impact parameter, $b$ & $0.47^{+0.34}_{-0.32}$ & $0.51^{+0.16}_{-0.23}$ \\
Time of Transit $t_{t}$~[$\rm BJD_{\rm TDB}$] & $2457749.1347^{+0.0035}_{-0.0032}$ & $2457740.0093^{+0.0044}_{-0.0041}$\\ 
Transit Duration $t_{14}$~[hours] & $3.98^{+0.21}_{-0.17}$ & $2.60^{+0.17}_{-0.15}$ \\ 
Equilibrium Temperature $T_{eq}$~ [K] & $750^{+170}_{-50}$ & $1425^{+79}_{-54}$ \\
$R_P$~[\rearth] & $2.45^{+0.35}_{-0.25}$ & $1.66^{+0.21}_{-0.19}$
 \\

\tablecomments{Note that $q_1$ and $q_2$ are the limb darkening coefficients as parameterized by \citet{kippingld}. }

\enddata
\end{deluxetable*}

\begin{figure*}[ht!]
    \centering
    \includegraphics[width=6in]{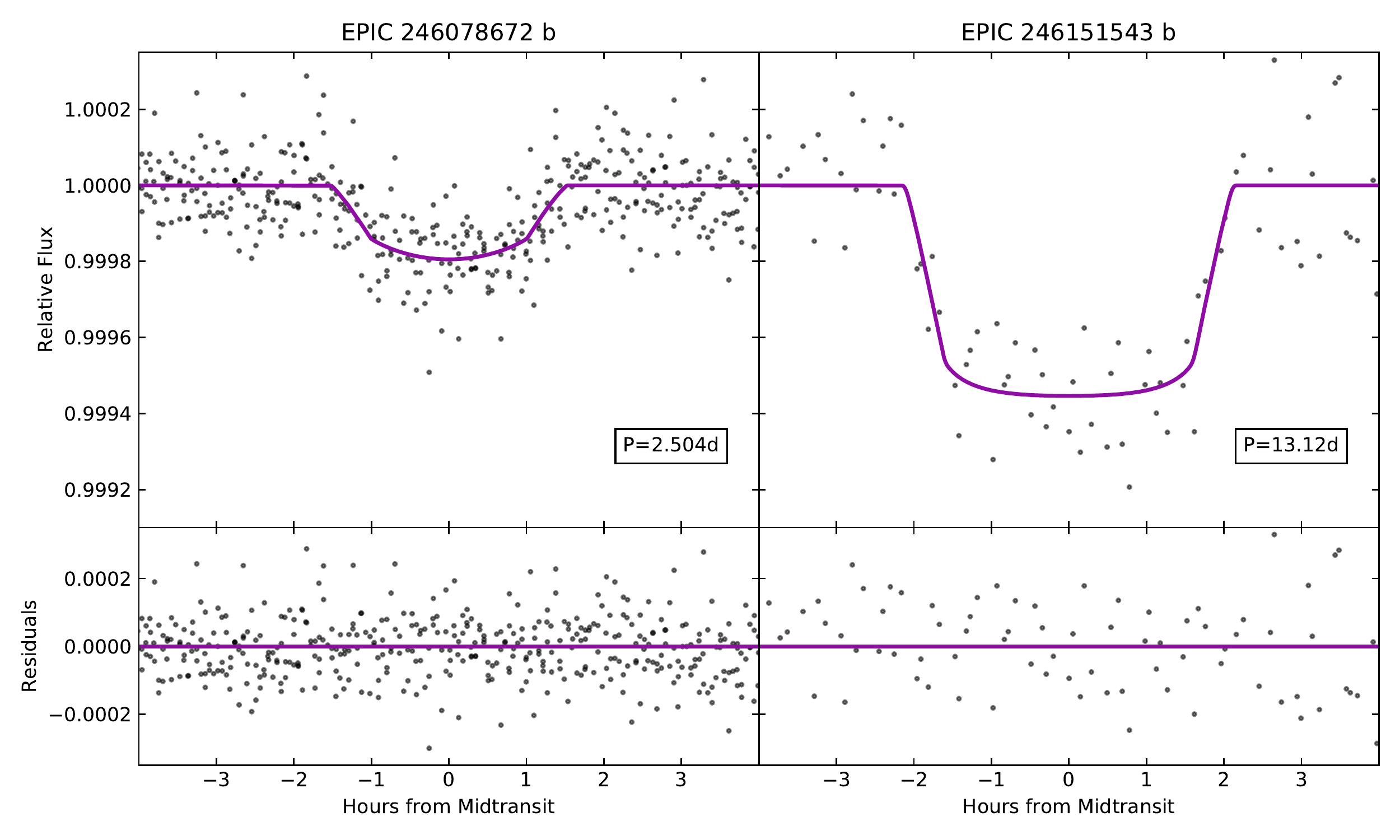}
    \caption{Transit light curves for the two new planets discovered: \thisplanettwo\ (left) and \thisplanetone\ (right). The light curves are shown with low-frequency stellar variability removed and folded on the orbital phase of the transits. The best-fit transit models are shown as purple solid lines.}
    \label{fig:newplanets}
\end{figure*}
\subsubsection{Statistical Validation}
\label{validation}

We calculated the false positive probabilities for the two newly identified planet candidates around \thisstarone\ and \thisstartwo\ using the open source \texttt{vespa} software \citep{vespa}, which implements the methodology of \citet{morton12} and \citet{morton16}. Given inputs describing the properties of a given candidate transit event, host star, and stellar neighborhood, \texttt{vespa} performs a model comparison between scenarios in which the transit signal is genuinely caused by a transiting exoplanet, and scenarios in which the transit signal is caused by some sort of astrophysical false positive scenario. We ran \texttt{vespa} with the transit light curves, spectroscopic parameters, 2MASS apparent magnitudes, \textit{Gaia} parallax, and constraints from speckle imaging as inputs, and found the false positive probabilities of both planets were quite low ($< 10^{-3}$). We therefore consider \thisplanetone\ and \thisplanettwo\ to be validated exoplanets. 

\section{Discussion}
\label{discussion}

\subsection{Newly Discovered Planets}

One result of our paper is that we identified two new exoplanets in K2 data using our averaged neural network model. Both planets are super-Earths orbiting G-dwarf stars; we list the stellar and planetary parameters for these two systems in Table \ref{bigtable}. 

\citet{rogers} showed that planets with radii $R\gtrsim1.6R_{\oplus}$ and orbital periods shorter than $\approx$ 50 days are usually not rocky. This describes \thisplanetone, which with a radius of $R = 2.45 R_{\oplus}$ is probably a ``puffy'' planet with a volatile envelope. This planet has an orbital period of 13.1 days, so it is strongly irradiated by its host star, but not so much that its volatile layer would be subject to photoevaporation. \thisplanettwo, with radius $R = 1.66 R_{\oplus}$, is right at the transition radius described by \citet{rogers}. This planet is likely still rocky because it is probably too close to its host star to have a hydrogen/helium envelope. Given its very short orbital period of 2.5 days, \thisplanettwo\ is decidedly not Earth-like; instead it is heated by its host star to scorching temperatures. The small radii and short orbital periods of \thisplanetone\ and \thisplanettwo\ are typical of the population of planets discovered by K2 \citep{mayo}.


While these planets themselves do not orbit particularly bright stars and are not particularly exciting for follow-up studies, our work is a proof of concept. In particular, if a similar study were to be performed on data from NASA's recently commissioned \textit{Transiting Exoplanet Survey Satellite} (TESS), which observes typically brighter stars than \Kepler\ or K2, we could potentially find highly valuable, previously missed planets well-suited for follow-up observations such as RV measurements, and once the \textit{James Webb Space Telescope} launches, spectroscopic observations of planetary atmospheres. TESS isn't as sensitive to small planets as \Kepler\ was, such as the planets in this paper, but it will be able to find many other, larger, planets for follow-up work. While TESS may not find as many super-Earths as \Kepler\ did, the ones it will find will be around brighter stars \citep[e.g.][]{huang2018}. 

\subsection{The continued value of human vetting}

Our model was built to recognize specific patterns in data and has only limited ability to recognize patterns beyond those it was trained to detect. This means that there may be unusual planetary systems out there that \textit{are} interesting but would not be recognized by our model and would not be flagged as interesting or worthy of study.

\begin{figure*}[ht!]
    \centering
    \includegraphics[width=6in]{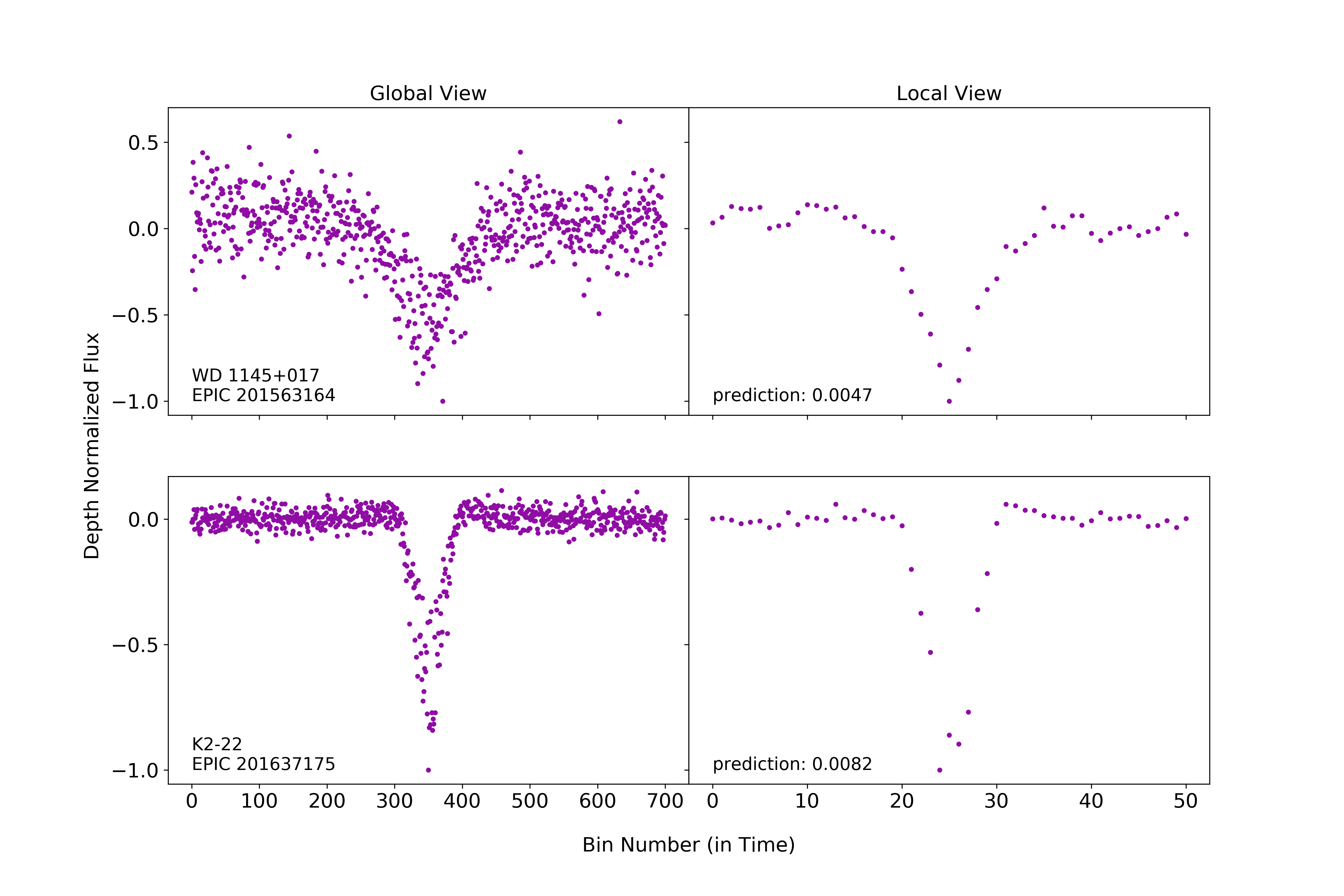}
    \caption{Two real planetary objects our model did not identify as likely planet candidates: WD 1145+017 b, a disintegrating planet orbiting a white dwarf star, and K2-22 b, a disintegrating planet orbiting an M-dwarf. These low predictions show that while the model is good at throwing out what obviously isn't a standard transit, there is still use in looking at TCEs by eye to identify some of the more unusual transit signals.}
    \label{fig:glob_loc_special}
\end{figure*}

For example, consider the two disintegrating planets discovered by K2: K2-22 b \citep{k2-22} and WD 1145+017 b \citep{wd1145}. These two signals were both detected by our transit search, though we excluded them from our training/validation/test set (see Section~\ref{training set}). After training our model to identify planet candidates and false positives, without including K2-22 b and WD 1145+017 b in the training set, we asked our model to classify the signals of these two disintegrating objects. Figure~\ref{fig:glob_loc_special} shows the global and local views of these two transit signals, as well as their predictions given by the model. The model assigns quite low predictions for the likelihood of these two signals being planets because the light curves are not typical for what a planet candidate looks like. The global views of the transits are not ``clean'' because the transit depth changes due to the disintegration. 

{\ron Another class of planets that may be misidentified by our model would be those showing significant transit timing variations (TTVs). It is more rare to see systems showing significant TTVs in K2 data compared to \Kepler, because K2's observational baseline is shorter. Over an 80 day K2 observing baseline, most TTVs can be approximated with a linear ephemeris and are therefore not noticeable in a phase-folded light curve. Nevertheless, a handful of systems identified by K2 do show significant TTVs.  We examined one such system, K2-146 \citep{hirano}, that was observed three times, in C5, C16, and C18. We tested our neural network on K2-146 b\footnote{We focus on K2-146 b, the transiting planet detected by \citet{hirano}, even though the signal of a second transiting planet (K2-146 c) became detectable in later campaigns.} using the data for all three campaigns separately. We found that during Campaign 5 and Campaign 18, the transit times could be fairly-well approximated with a linear ephemeris, so the phase-folded light curve seen by our neural network looked like a ``normal'' transiting planet without TTVs, and the model gave the TCEs fairly high predictions of 0.45 and 0.28, respectively. In Campaign 16, however, the TTVs were not as well approximated with a linear ephemeris, leading to a distorted phase-folded light curve and a much lower prediction of 0.0014. }

To a human, these {\ron atypical} signals appear interesting and require further analysis, but the neural network only identifies ``typical'' signals as planets. A method such as this is good for picking out typical, standard transiting planets, but may pass over special or interesting ones.

The fact that our neural network did not identify the two disintegrating planets WD 1145+017 b and K2-22 b and the TTV system K2-146 as planet candidates shows the enduring value of humans classifying planet candidates by eye. Humans are good at recognizing unusual signals that machines will mis-classify or not recognize as interesting, which is crucial for discovering interesting and odd facets of the universe.

\subsection{Improvements/Future Work}

While our neural network was successful at identifying new planet candidates, there is room for some significant improvements. We list a handful of possible future avenues of work to improve the model here. 

\begin{enumerate}
    \item Improved training set {\ron labels}. Currently, our training set is composed of thousands of targets that were labeled by the human eye. Most TCEs in our training set were each scrutinized for only about a second  during triage by only one person, and are therefore subject to some error. There are almost certainly incorrectly labeled signals in the training set. Moreover, the triage step was performed on the TCEs as the data for each campaign became available over the course of four years of the K2 mission (2014-2018). Undoubtedly, our criteria for assigning training set labels in triage changed somewhat over that time period as we learned more about the characteristics of K2 data and as our scientific goals evolved. Improving the accuracy of the training set labels will allow the model to learn more efficiently and correctly, and will make it more accurate to evaluate the success of the model by comparison with the test set. 
    \item {\ron We would like to extend our training set with simulated planet transits and false positives, but this approach presents a few challenges. First, artificial signals could bias our training set if added in the wrong proportions (e.g. too many planets vs false positives, or incorrect distribution of orbital period), and we might not know the correct proportions to simulate. Second, we might not be able to accurately simulate all types of instrumental and astrophysical false positives. Inverted and scrambled light curves have previously been used to produce artificial false positives \citep{dr25-kp-efficiency,dr25-rv-efficiency}, but if these signals are sufficiently different from real false positives then including them in our training set would not be useful and might make us overestimate our model's performance. Future work could investigate how well these simulated false positives match the population of real false positives, or explore additional ways to accurately simulate false positive signals.}
    \item Our data representation could benefit from improving the interpolation method we use when preparing the global and local views of each light curve. Currently, when there are bins in either view without data points, we {\ron compute estimated values for those bins by linearly interpolating between neighboring bins} (see Section~\ref{preprocessing} for further details). This type of interpolation can sometimes disguise the true nature of the TCE, perhaps by removing periodic signals or sudden jumps in the data, which hinders the model's ability to learn from and judge those signals. A different, more sophisticated method for interpolating data when preparing the local and global views may improve the model's performance. {\ron Alternatively, we could  avoid interpolation altogether by using a more sophisticated input representation. For example, we could include an additional input value for each bin to indicate whether the bin was empty, which would allow the model to learn how to deal with missing data itself. Note that we cannot simply delete missing bins, since that would distort the time axis and potentially change the shape of a transit.}
    \item More information and features could also be provided to the model to improve its ability to distinguish false positives from planet candidates. Currently, our model is forced to make decisions with only two views of the phase-folded and binned light curve as inputs. But in the triage and vetting processes (Sections~\ref{triage} and \ref{vetting}) we had access to more information (see Figures \ref{fig:triage}, \ref{fig:vetting1}, and \ref{fig:vetting2}) than just the phase-folded light curves to classify each signal, including the un-phase folded light curve, the spacecraft's position during transits (which can identify instrumental systematics), the behavior of the transit signals in light curves produced with different photometric apertures (which can identify background contamination), and differences between even-numbered and odd-numbered transits (which can identify eclipsing binaries). Giving our neural network more features like those used during human vetting would likely make its classifications more accurate \citep[for example, see][]{ansdell}. 
\end{enumerate}



Our work is also highly applicable to other data sets, such as TESS. Like \Kepler\ and K2, TESS will detect thousands of TCEs in need of classification. Modifying our neural network to classify TCEs from TESS will face similar challenges as we did going from Kepler to K2 because of the shorter time span of TESS light curves compared to even K2. TESS's larger pixels will pose additional challenges by introducing more false positive signals, and may benefit even more from information about image centroids. We are currently testing strategies for adapting our neural network to classify TCEs from TESS data (L.~Yu~et~al.,~in~prep). A training set of signals from TESS as large as the one used here may take several years to build up, but a technique such as domain adaptation \citep[like has been used in galaxy classification;][]{domainadaptationdessdss} or partially synthetic training sets may be used to speed up the process.

Future work is also needed to characterize our model's performance on the underlying distribution of all TCEs. Our training and test sets necessarily differ from this distribution because we did not have access to ground-truth labels for all TCEs in our labeling process.
For example, we used a ``planet candidate until proven otherwise'' rule and also discarded any TCEs whose true classification was ambiguous (see Section~\ref{vetting}), which gave us a modified distribution of labeled TCEs. If we had adopted an alternative labeling procedure, such as ``false positive until proven otherwise,'' we would have obtained a different modified distribution. This observation has two implications: (i) since we trained our model on the modified distribution, it may perform worse on the underlying distribution, and (ii) since we tested our model on the modified distribution, the metrics we measured over our test set might not reflect its performance on the underlying distribution. Unfortunately, it is challenging to evaluate the performance of \textit{any} TCE classifier (e.g. human, Robovetter, neural network) on the underlying distribution without having access to ground-truth labels for all TCEs. Prior work has used artificially injected planet signals to estimate the true positive rate and simulated false positives to estimate the false negative rate \citep{thompson2018,dr25-kp-efficiency,dr25-rv-efficiency}, but this approach is limited by our ability to identify all types of false positives and accurately simulate them in the right proportions. Better ways to evaluate the performance of TCE classifiers will be crucial for improving estimates of planetary occurrence rates from \Kepler\ data, regardless of the classification method used.

\subsection{Towards Occurrence Rates in K2 Data}


Our model is not yet ready to measure planetary occurrence rates in K2 data, but we hope that the work outlined in the previous section will eventually make this possible. First, we must be able to characterize our model's performance on the underlying distribution of all TCEs, not just the modified distribution of TCEs in our test set. Second, we must improve our model to make it more accurate at rejecting false positive signals.
Currently, if we select a classification threshold so that our model recalls 90\% of planet candidates in our test set, we only have a precision (or purity of the planet candidate sample) of around 50\%. 
As a goal, we estimate that an improved neural network, able to measure occurrence rates, would ideally have a recall of at least 90\% and a precision greater than 95\% in order to introduce errors on planet occurrence rates less than 5\% while remaining sensitive to most planet candidates in the sample. Since the K2 data consists of approximately 95\% false positives, our model would then have to correctly classify planets $90\%$ of the time and false positives $99.75\%$ of the time.

Since K2 looked at fields all across the ecliptic plane, occurrence rates based on K2 would allow us to compare specific characteristics of different planet populations. K2 has campaigns looking at different galactic longitudes and distances from the galactic midplane, which correspond to both a difference in age and a difference in metallicity. An improved neural network could enable accurate measurements of the abundances of planets in those regions to improve understanding on how age and metallicity impact planet formation. K2 also observed regions of the galaxy that were densely packed with stars, which could impact the way that planets form around their host stars. Occurrence rates in K2 data might help us understand how and why planet formation and evolution is dependent on galactic birthplace. 

\section{Summary}

In this work, we have trained a neural network, which we call \texttt{AstroNet-K2}, to identify planet candidates in K2 data for the first time. Previously, neural networks have been used to classify planet candidates in data from the original \Kepler\ mission, and have the promise of helping to deliver more accurate planet occurrence rates. Here, we extended this work on \Kepler\ data to data from the K2 mission. Our results can be summarized as follows: 

\begin{enumerate}
    \item We have built and trained a neural network to classify possible planet signals detected in K2 data into likely planet candidates and false positives. Our neural network is quite successful - it is able to correctly classify signals with an accuracy of 98\% and an AUC of 0.988 on our test set. While the performance of our network is not quite at the level required to generate fully automatic and uniform planet candidate catalogs, it serves as a proof of concept.  
    \item We used our neural network to rapidly search through a large set of previously un-classified possible planet signals. It identified several signals that had previously slipped through the cracks in our planet detection pipeline, and that we had not previously noticed as likely planet candidates. We performed follow-up observations of two of these candidates, \thisplanetone\ and \thisplanettwo, and were able to validate these objects as genuine exoplanets. The new planets are both in close orbits around their host stars and are intermediate to the Earth and Neptune in size. 
    \item This work can be applied other datasets in the future. In particular, the recently commissioned TESS mission is currently observing large swaths of sky in the Southern ecliptic hemisphere, and detecting large number of TCEs. Like K2, TESS will observe stars for shorter periods of time than the original \Kepler\ mission, so the modifications we have made to this network for K2's shorter time baseline will be useful for adapting to TESS. 
\end{enumerate}

Developing a neural network to rapidly and automatically classify planet candidates from the K2 mission is a critical step on the path towards using K2 to determine planet occurrence rates, and comparing the populations of exoplanets in different galactic environments. Previously, all planet candidate catalogs from K2 data have relied on humans to identify planet candidate signals, whose biases and inconsistencies introduce uncertainties in the ultimate occurrence rate measurements. The uniform predictions of a neural network may eventually make it possible to measure these rates more accurately.

\label{summary}

\acknowledgments
AD acknowledges support from the John W. Cox Endowment for the Advanced Studies in Astronomy. AV's work was performed under contract with the California Institute of Technology (Caltech)/Jet Propulsion Laboratory (JPL) funded by NASA through the Sagan Fellowship Program executed by the NASA Exoplanet Science Institute. 

This research has made use of NASA's Astrophysics Data System and the NASA Exoplanet Archive, which is operated by the California Institute of Technology, under contract with the National Aeronautics and Space Administration under the Exoplanet Exploration Program.

This paper includes data collected by the \Kepler\ mission. Funding for the \Kepler\ mission is provided by the NASA Science Mission directorate. Some of the data presented in this paper were obtained from the Mikulski Archive for Space Telescopes (MAST). STScI is operated by the Association of Universities for Research in Astronomy, Inc., under NASA contract NAS5--26555. Support for MAST for non--HST data is provided by the NASA Office of Space Science via grant NNX13AC07G and by other grants and contracts. This work has made use of data from the European Space Agency (ESA) mission {\it Gaia} (\url{https://www.cosmos.esa.int/gaia}), processed by the {\it Gaia} Data Processing and Analysis Consortium (DPAC, \url{https://www.cosmos.esa.int/web/gaia/dpac/consortium}). Funding for the DPAC has been provided by national institutions, in particular the institutions participating in the {\it Gaia} Multilateral Agreement.

We wish to recognize and acknowledge the very significant cultural role and reverence that the summit of Maunakea has always had within the indigenous Hawaiian community.  We are most fortunate to have the opportunity to conduct observations from this mountain.

Facilities: \facility{Kepler/K2, FLWO:1.5m (TRES), Gemini:Gillett (\Alopeke)}

{\ron Software: \facility{TensorFlow \citep{abadi2016tensorflow}, AstroNet \citep{shallue}, Numpy \citep{np}, Scipy \citep{scipy}, Matplotlib \citep{plt}}}


\input{table.tex}

\clearpage

\end{document}

%% file: table.tex
\clearpage
\LongTables
\begin{deluxetable*}{ccccccccc}
\tablewidth{0pt}
\tabletypesize{\scriptsize}
\tablecaption{\label{tab:stellar} K2 Planet Candidate Training/Validation/Test Set}
\tablehead{
  \colhead{EPIC} &
  \colhead{Planet} &
  \colhead{Period} &
  \colhead{Time of Transit} &
  \colhead{Duration} &
  \colhead{Training} &
  \colhead{Campaign} &
  \colhead{$R_p/R_\star$} &
  \colhead{Impact} \\
  \colhead{ID} &
  \colhead{Number} &
  \colhead{(days)} &
  \colhead{(BJD-2454833)} &
  \colhead{(hours)} &
  \colhead{Label} &
  \colhead{Number} &
  \colhead{} &
  \colhead{Parameter} 
}
\startdata
200001049 & 1 & 28.94461125 & 2002.122414 & 2.990041376 & J & 1 & 2.555702219 & 1.248386684  \\
201124136 & 1 & 35.38110738 & 1990.254887 & 2.939212598 & J & 1 & 0.162759219 & 1.046836871  \\
201126503 & 1 & 1.194869833 & 1977.375616 & 1.7932546   & C & 1 & 0.061295114 & 0.775590604  \\
201131821 & 1 & 10.5507237  & 1987.073905 & 1.464041586 & J & 1 & 0.099763623 & 0.517321219  \\
201133060 & 1 & 7.843923319 & 1978.523391 & 1.591163347 & J & 1 & 0.089307523 & 0.000494194 \\
201136242 & 1 & 0.245037064 & 1977.338043 & 0.141141349 & J & 1 & 0.068636433 & 0.01         \\
201137589 & 1 & 0.12262433  & 1977.287719 & 0.070631614 & J & 1 & 0.065830297 & 0.01         \\
201141876 & 1 & 0.163459033 & 1977.30514  & 0.094152403 & J & 1 & 0.338031438 & 0.01         \\
201143107 & 1 & 7.843922992 & 1978.52283  & 1.564043424 & J & 1 & 0.067069183 & 0.000499999 \\
201145205 & 1 & 28.402966   & 1977.380815 & 3.105401341 & J & 1 & 0.323971051 & 1.095744208  \\
201145621 & 1 & 28.40594082 & 1977.366573 & 0.991484362 & J & 1 & 0.086576647 & 0.376126662  \\
201146489 & 1 & 21.42846654 & 1989.586459 & 6.004301954 & E & 1 & 0.128439312 & 1.028751044 \\
201148067 & 1 & 40.01727508 & 1977.405056 & 5.361641493 & J & 1 & 0.130607363 & 0.840070237  \\
201148768 & 1 & 38.99426968 & 1986.589685 & 1.524765554 & J & 1 & 0.1443235   & 0.00056013  \\
201149979 & 1 & 40.03607983 & 1977.403191 & 8.748567385 & J & 1 & 0.386937487 & 1.191895831  \\
201150341 & 1 & 28.96556197 & 1986.583244 & 1.140516447 & J & 1 & 0.107478013 & 0.235029835 \\
201150403 & 1 & 35.0347551  & 1990.547456 & 2.483563041 & J & 1 & 0.072085604 & 0.001441808  \\
201150433 & 1 & 25.39554322 & 1990.549255 & 1.598189432 & J & 1 & 0.076518442 & 0.451982693  \\
201150492 & 1 & 0.281400719 & 1977.335881 & 0.862632823 & J & 1 & 0.229233474 & 0.057482799  \\
\enddata
\tablecomments{The training set is labeled either as J (which stands for junk, and typically are either instrumental systematics or stellar variability), C (which stands for planet candidate), or E (which stands for eclipsing binary star). The full version of this table is available in the LaTeX package on ArXiv. }

\end{deluxetable*}